\documentclass[prb,aps,twocolumn,showpacs,preprintnumbers,superscriptaddress,amsmath,amssymb]{revtex4-1}
\usepackage{graphicx}% Include figure files
\usepackage{dcolumn}% Align table columns on decimal point
\usepackage{bm}% bold math
\usepackage{color}

\begin{document}

\title{Correlation strength and orbital differentiation across the phase diagram of plutonium metal}

\author{W. H. Brito}
\affiliation{Departamento de  F\'{\i}sica, Universidade  Federal de Minas Gerais, C. P. 702, 30123-970, Belo Horizonte, MG, Brazil.}
\affiliation{Department of Physics and Astronomy, Rutgers University, Piscataway, New Jersey 08854, USA.}
\author{G. Kotliar}
\affiliation{Department of Physics and Astronomy, Rutgers University, Piscataway, New Jersey 08854, USA.}
\affiliation{Condensed Matter Physics and Materials Science Department, Brookhaven National Laboratory, Upton, New York 11973, USA.}

\begin{abstract}
We  compare the trends on the strength of electronic correlations across the different phases of elemental Pu focusing on its site and orbital dependence, using a combination of density functional theory (DFT) and dynamical mean field theory (DMFT) calculations within the vertex corrected one crossing approximation. We find that Pu-5$f$ states are more correlated in $\delta$-Pu, followed by some crystallographic sites in $\alpha$ and $\beta$ phases. In addition, we observe that Pu-5$f_{5/2}$ and Pu-5$f_{7/2}$ orbital differentiation is a general feature of this material, as is site differentiation in the low symmetry phases. The Pu-5$f_{5/2}$ states show Fermi liquid like behavior whereas the Pu-5$f_{7/2}$ states remaining incoherent down to very low temperatures. We correlate the correlation strength in the different phases to their structure and the Pu-5$f$ occupancy of their crystallographic sites. 
\end{abstract}

\maketitle

\section{Introduction}

Plutonium (Pu), the sixth member of the actinide series, has been referred to by Siegfried Hecker as “an element at odds with itself”,  and  “a  physicist dream  and metallurgist nightmare”.~\cite{LANL_Pu,hecker_pms} This is because the solid state properties of this material,  have been exceptionally challenging to describe or understand from a theoretical perspective  as it behaves very differently from all elemental solids.  For example, elemental Pu exhibits six structural phases between 100 K and 756 K,~\cite{moore_review} more phases than any elemental solid in such a narrow temperature range. In contrast to simple metals,  metallic Pu crystallizes in  monoclinic low-symmetry structures at low temperatures,  known as $\alpha$ and $\beta$ phases.  Upon heating the $\beta$ phase undergoes a structural transition to the $\gamma$ phase, which is a face centered orthorombic structure, followed by $\delta$, $\delta'$, and $\epsilon$ phases.~\cite{moore_review} The $\delta$ phase has been the subjected of most theoretical studies,  and was regarded as its most exotic properties, such as  the ``missing magnetism'',~\cite{lashley,heffnerPRB} large specific heat,~\cite{lashleyPRL,havelaPRB} and negative thermal expansion which is also observed in $\delta'$ phase.~\cite{moore_review}

Density functional theory (DFT), and its extensions to magnetic systems such as spin density functional theory target ground state properties. It has been applied to study the energetics of metallic Pu since the early nineties. Initially it was viewed that the low symmetry more dense phases of Pu, were well described within the standard exchange correlation potentials such as the generalized gradient approximation (GGA), while the high symmetry phases, in particular the $\delta$ phase, required treatments that go beyond.~\cite{erickson_wills_las}Moreover, it was known at that time that non magnetic DFT treatments  were unable to describe the equilibrium volume of Pu.~\cite{dftpu2,dftpu1} At their experimental equilibrium volumes, $\alpha$ and $\delta$-Pu were predicted to have an antiferromagnetic ground state.~\cite{kutepov_pu} Later, S\"{o}derlind and Sadigh obtained equilibrium volumes and bulk moduli of all six phases of elemental Pu  in good agreement with experimental data, although their calculations also predict the existence of large local moments in all phases.~\cite{soderlindPRL} More recently, the lattice dynamics and elastic constants for the $\delta$ and $\epsilon$ phases have been obtained within DFT methods,~\cite{soderlind_ConstPRB,soderling_phonDelta,soderling_phonEps} where disordered magnetic moments were considered. DFT+U calculations were also used to describe the structural properties of elemental Pu, though by using small $U$ values ($U \sim$ 1 eV),~\cite{amadon_JPhys,qiu_CMS} which also predict a magnetically ordered ground state. For recent reviews see Refs.~\onlinecite{soderlind_Adv,review_soderlind, kutepov_appsci}. 
Hence  static mean field theories, can reproduce energetics and structural properties, while predicting some form of magnetism, in the form of ordered or disordered local moments.~\cite{savrasov_ldau_PRL} 

Dynamical mean field theory (DMFT) in combination with electronic structure methods, is a technique, that in principle can determine spectra and energies, but it is much more computationally expensive, and it has been used by many groups to investigate elemental Pu.~\cite{elihu,xidai_Science,shimOCAPu,marianetti_PRL,willsNatCom,janoschek,amadon_PRB,whb_betapu,zhu_holepocket,LHuang_Pu} An early success of this approach, using relatively simple impurity solvers, was to account for the correct lattice constants of $\delta$-Pu in a non-magnetic framework and predict its phonon spectra,~\cite{xidai_Science}  which was later measured in inelastic x-ray scattering experiments.~\cite{wong_Science} 
An important issue within DMFT is the value of the Pu-5$f$ occupancy, which in most studies is slightly bigger than $5$, but the precise $f$ occupancy does depend on the form of the double counting corrections used.~\cite{zhu_prb} Constraints on this occupancy come from x-ray absorption measurements.~\cite{tobin_xas} From a broader perspective DMFT studies served to established the relevance of concepts connected with the delocalization-localization transition,~\cite{johansson_pmg} placing $\delta$-Pu close to the border between itinerant and localized behavior, but now recognizing that {\it all} phases of Pu are correlated, even the $\alpha$ phase, which has a much smaller volume, still has correlation features in its spectra. 

Marianetti et al.~\cite{marianetti_PRL}, showed that at high temperatures and for large volumes $\delta$-Pu has a Pauli-like magnetic susceptibility, and the moments are screened at a coherence temperature of around 800 K. Further DMFT calculations pointed out that $\delta$-Pu is a mixed valent metal.~\cite{shimOCAPu} Magnetically ordered states of elemental actinides were also explored within DMFT, where it was reported that magnetism is stable in curium but not in plutonium,~\cite{shimOCAPu} which was then firmly established to be paramagnetic. These studies explained the Pu ``missing  magnetism'' and other anomalies through a mixed valence mechanism, where fluctuations between the $f^5$ and $f^6$ configurations are important. This picture was later confirmed experimentally by inelastic neutron scattering where a resonance was observed at finite frequency.~\cite{janoschek}

Photoemission and spectral function play a very important role in the theory of actinides. Dynamical mean field theory provides several concepts for their interpretation. It was shown that multiplet splittings, rather than the hybridization, determine the width of the Hubbard bands in materials such as americium.~\cite{savrasov_americium} DMFT histograms were developed to provide an underpinning to the concepts of valence in the solid state as they allow to visualize the probability of finding a site at different atomic configurations.~\cite{shimOCAPu} Successful comparisons with photoemission experiments were carried out in several Pu compounds where quasiparticle multiplets were shown to be a signature of mixed valence.~\cite{pusi_multiplets,chuckpu} Angle resolve studies are now possible and recently Fermi surface pockets in $\delta$-Pu, were predicted using DMFT.~\cite{zhu_holepocket}

More recent DMFT work have begun to address the electronic properties of the other phases of Pu besides the $\delta$ phase. In the work of Zhu et al.~\cite{willsNatCom} the authors found that in $\alpha$-Pu the correlation strength is very site dependent. This observation can account for why the resistivity  and susceptibility in this phase is larger than in the $\delta$ phase, even though $\alpha$ has a smaller volume. Rotationally invariant slave boson (RISB) and the Gutzwiller approximation,  can be also viewed as simplified embedding methods in the spirit of DMFT. The evolution of the occupancies and the energetics of all the Pu phases were successfully addressed in Ref.~\onlinecite{lanata_risb}. Site-dependent correlations were also reported for $\beta$-Pu.~\cite{whb_betapu} in qualitative agreement with the experimental side results by L. Havela et al.~\cite{havela_puos}.

In particular, it was reported that different orbitals in the same site exhibit different levels of correlations, {\it i.e.} orbital differentiation. This is a very general phenomena, which has been studied extensively in the context of Hund metals~\cite{Haule_Hunds,ziping,medici_PRB1,medici_PRL_2011,lanata_iron,fabian_PRB} and Mott systems.~\cite{anisimov_ruth, kogaPRL}. More recent work,~\cite{LHuang_Pu} considered the spectra across all the phases of Pu, using the continuous time quantum Monte Carlo (CTQMC) solver, but simplified their calculations to have a single representative site in all the phases. They  identified quasiparticle multiplets in the  spectra of  the $\alpha$, $\beta$, and $\gamma$ phases, which are thus also  Racah metals.

In this work, we continue to address the broad picture of how correlations evolve across all the phases of Pu using DFT+DMFT(OCA). Our work is complementary to Ref.~\onlinecite{LHuang_Pu} as we focus on the site dependence of the spectral functions which we show to be important and we do not rely on analytic continuation.   
We explore the degree of correlations and Pu-5$f$ orbital differentiation in all the six phases. According to our calculations, the Pu-5$f_{5/2}$ states are more correlated in $\delta$-Pu followed by some sites in $\alpha$ and $\beta$ phases. For all phases the Pu-5$f$ spectral function is composed of quasiparticle peaks near the Fermi energy,  Hubbard bands at higher energies, and less intense quasiparticle multiplets which are characteristic of Hund's-Racah metals.~\cite{shick_racah,Lichtenstein_Racah,whb_betapu} Overall, the Pu-5$f_{5/2}$ and Pu-5$f_{7/2}$ spectral functions are very distinct, with Pu-5$f_{5/2}$ electrons in a Fermi-liquid like regime whereas electrons in the Pu-5$f_{7/2}$ states are very incoherent. This feature is captured by the renormalized imaginary part of Pu-5$f$ self-energies, which explains the common feature of orbital differentiation throughout Pu phase diagram.

\section{Computational Methods}
\label{method}

Our calculations were performed using the fully charge self-consistent DFT+embedded-DMFT approximation.~\cite{hauleWK} 
The DFT part of our calculations  were carried out within the full potential linearized augmented plane wave (FP-LAPW) method  and  Perdew-Burke-Ernzehof generalized gradient approximation (PBE-GGA),~\cite{pbe} as implemented in Wien2K package.~\cite{wien} 
We mention that the distinct crystallographic sites of the low-symmetry structures gives rise to a  several DMFT quantum impurity problems, which were solved self-consistently.  
The DMFT effective impurity problems were solved using the vertex-corrected one-crossing approximation (OCA)~\cite{pruschke} with the on-site Coulomb repulsion $U = 4.5 $ eV and Hund's coupling $J = 0.512$ eV.  In our previous work on $\beta$-Pu~\cite{whb_betapu} we showed that OCA can captures the essential features of elemental Pu in comparison with more precise and computationally expensive continuous time quantum Monte Carlo (CTQMC) calculations.

The values of $U$ and $J$ used in this work are in conformity with previous calculations which described successfully to the valence-fluctuating ground state of $\delta$-Pu within our implementation,~\cite{janoschek} which in turn takes into account all the itinerant and correlated states within a 20 eV energy window around the chemical potential.  We mention that different implementations used small U values to study elemental Pu.~\cite{amadon_PRB}
Finally, we used the standard fully localized-limit form~\cite{anisimovEdc} for the double-counting correction term, with $n_{f}^{0}=5.2$, which is the average occupation of Pu-5$f$ states in the $\delta$-Pu.~\cite{shimOCAPu}

\section{Results and Discussions}

\subsection{High-symmetry $\gamma$, $\delta$, $\delta'$, and $\epsilon$ phases}

\begin{figure}[!htb]
\includegraphics[scale=0.66]{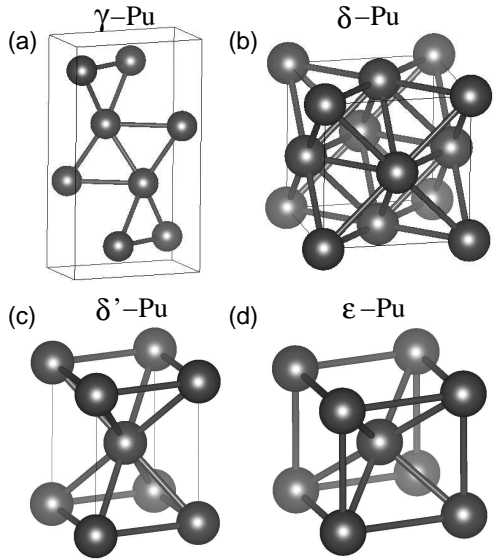}
 \caption{Crystal structures of (a) $\gamma$ (face centered orthorombic, space group $I4/mmm$) ,  (b) $\delta$ (face centered cubic, space group $Fm-3m$), (c) $\delta'$ (body centered tetragonal, space group $I4/mmm$ ), and (d) $\epsilon$ (body centered cubic, space group $Im-3m$) phases.   Pu atoms are represented as black spheres.}
\label{fig:crystal_HS}
\end{figure}

From around 200 K up to its melting point, Pu has four high-symmetry crystallographic phases, namely $\gamma$, $\delta$, $\delta'$, and $\epsilon$ (see Fig.~\ref{fig:crystal_HS}). The former has a face centered orthorombic structure, with two atoms within the unit cell, followed by the $\delta$ phase which is a face centered cubic structure. $\delta'$-Pu, in turn, has a body centered tetragonal structure while the highest temperature solid phase $\epsilon$ is a body centered cubic.

\begin{figure}[!htb]
 \includegraphics[scale=0.4]{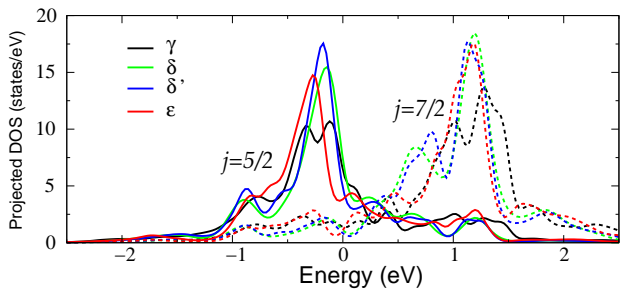}
 \caption{DFT(GGA) projected density of states for $\gamma$ (black), $\delta$ (green), $\delta'$ (blue), and $\epsilon$  (red) phases.  Continuous (dashed) lines denote the Pu-5$f_{5/2}$ (Pu-5$f_{7/2}$) projected density of states. }
 \label{fig:pdos_DFT_hs}
\end{figure}

Although the high-symmetry phases crystallize in different simple structures, the Pu-5$f$ states  have very similar features within all these phases. As shown in the DFT projected density of states (Fig.~\ref{fig:pdos_DFT_hs}), one can notice that the Pu-5$f_{5/2}$ ($j = 5/2$) states are located just below the Fermi energy while the Pu-5$f_{7/2}$ ($j=7/2$) are higher in energy, centered around 1.15 eV. There is also an orbital mixing which leads to a finite occupancy of the Pu-5$f_{7/2}$ states. The splitting of the Pu-5$f$ manifold occurs due to the strong Pu spin-orbit coupling, which is similar to crystal field splittings of $d$ manifold in materials with $d$ electrons.

\begin{figure}[!htb]
 \includegraphics[scale=0.4]{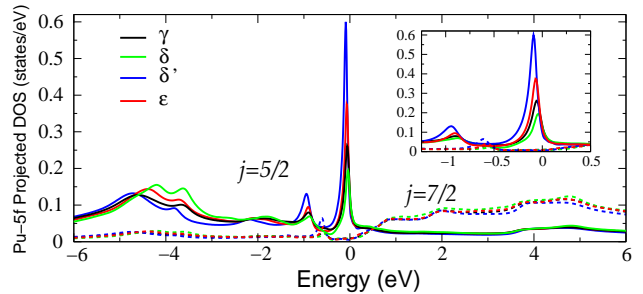}
 \caption{DFT+DMFT(OCA)  Pu-5$f$ projected density of states at 500 K for $\gamma$ (black), $\delta$ (green), $\delta'$ (blue), and $\epsilon$  (red).  Continuous (dashed) lines denote the Pu-5$f_{5/2}$ (Pu-5$f_{7/2}$) projected density of states. Insets display the projected density of states in a smaller energy window (-1.5 eV $\leq$ E$_{F}$ $ \leq$ 0.5 eV).}
 \label{fig:pdos_DMFT_hs_OCA}
\end{figure}

The effects of dynamic electronic correlations on the Pu-5$f$ excitation spectra can be observed in Fig.~\ref{fig:pdos_DMFT_hs_OCA}, where we show the DFT+DMFT Pu-5$f$ projected density of states. The fingerprint of strong electronic correlations in these materials is the presence of strongly renormalized quasiparticle peaks near the Fermi energy, Hubbard bands at higher energies, and quasiparticle multiplets in their excitation spectra.  It is important to mention that quasiparticle peaks appear due to antiferromagnetic interaction of the Pu-5$f$ electrons with the surrounding conduction electrons. These heavy quasiparticles states, which are of Pu-5$f_{5/2}$ character,  are more intense in $\delta'$ phase followed by $\epsilon$, $\gamma$ , and $\delta$ (see inset in Fig.~\ref{fig:pdos_DMFT_hs_OCA}). This indicates that the Kondo temperature is larger in $\delta'$ phase in relation to the other phases.
For all these phases we find the multiplets  around -0.9 eV and -0.6 eV, of Pu-5$f_{5/2}$ and Pu-5$f_{7/2}$ characters, respectively.
We mention that these quasiparticle multiplets are a common feature in the excitation spectra of Pu-based materials.~\cite{shimOCAPu,chuckpu,whbpu115, whb_betapu}
Another important finding is the structure independent Pu-5$f_{7/2}$ spectral function. 
As can be notice in Fig.~\ref{fig:pdos_DMFT_hs_OCA}, these states are essentially gapped at the Fermi energy and give rise to a upper Hubbard band around 4 eV.

\begin{figure}[!htb]
 \includegraphics[scale=0.4]{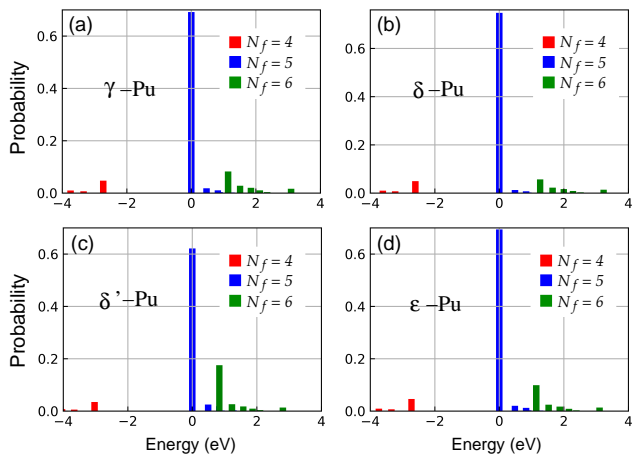}
 \caption{Valence histogram of Pu-5$f$ states  of (a) $\gamma$, (b) $\delta$, (c) $\delta'$, and (d)$\epsilon$-Pu at 500 K.  N$_f$ denotes the number of electrons in the Pu-5$f$ manifold.}
 \label{fig:histo_DMFT_hs_OCA}
\end{figure}

The fluctuations of the electron number within the Pu-5$f$ impurity states can be seen in the valence histograms shown in Fig.~\ref{fig:histo_DMFT_hs_OCA}.  Overall, the 5$f$ states in the high-symmetry phases have a mixed valence state,  where the configurations with N$_f$ = 5 (5$f^{5}$) and N$_{f}$ = 6 (5$f^{6}$), have the larger probabilities. These larger probabilities indicate that the Pu-5$f$ electrons spend a considerably fraction of time in the 5$f^{5}$ and 5$f^6$ configurations. These findings are in agreement with previous DMFT calculations for the $\delta$-Pu~\cite{shimOCAPu} and more recent DMFT(CTQMC) reported in Ref.\onlinecite{LHuang_Pu}.  We also find that the probabilities of these two configurations have some dependence on the crystallographic environment, as can be noticed for the 5$f^{5}$ configuration, which has the largest probability in $\delta$-Pu, whereas the 5$f^{6}$ is more probable in the $\delta'$ phase. 

\subsection{Low-symmetry $\alpha$ and $\beta$ phases}

\begin{figure}[!htb]
\includegraphics[scale=0.45]{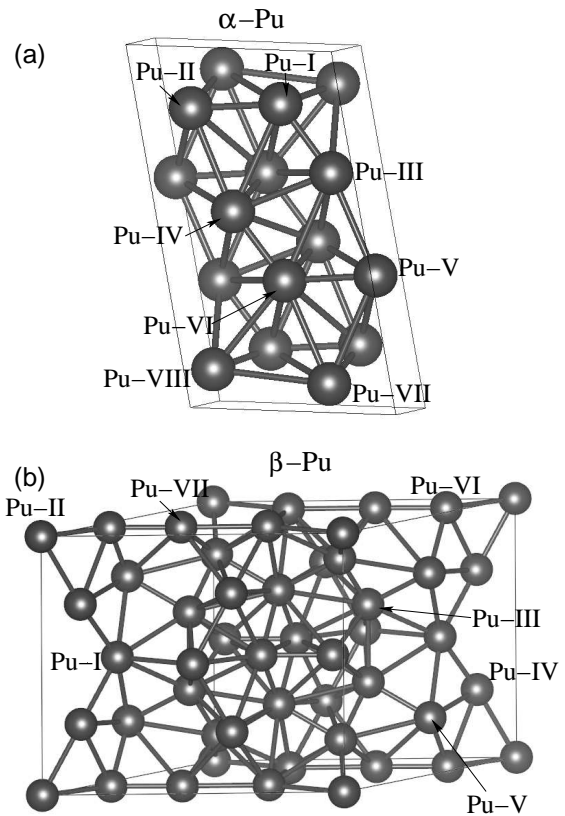}
 \caption{Crystal structures of (a)$\alpha$ (space group $P21/m$) and (b) $\beta$-Pu (space group $I2/m$), where the Pu atoms are represented as black spheres. The inequivalent Pu sites in each phase are schematically indicated by the arrows.}
\label{fig:struct_lowSymm}
\end{figure}

In Fig.~\ref{fig:struct_lowSymm} we present the  structures of the low-symmetry crystallographic phases, namely $\alpha$ and $\beta$ phases.  In $\alpha$-Pu ($\beta$-Pu) there are eight (seven)  inequivalent crystallographic sites.

\begin{figure*}[!htb]
\includegraphics[scale=0.54]{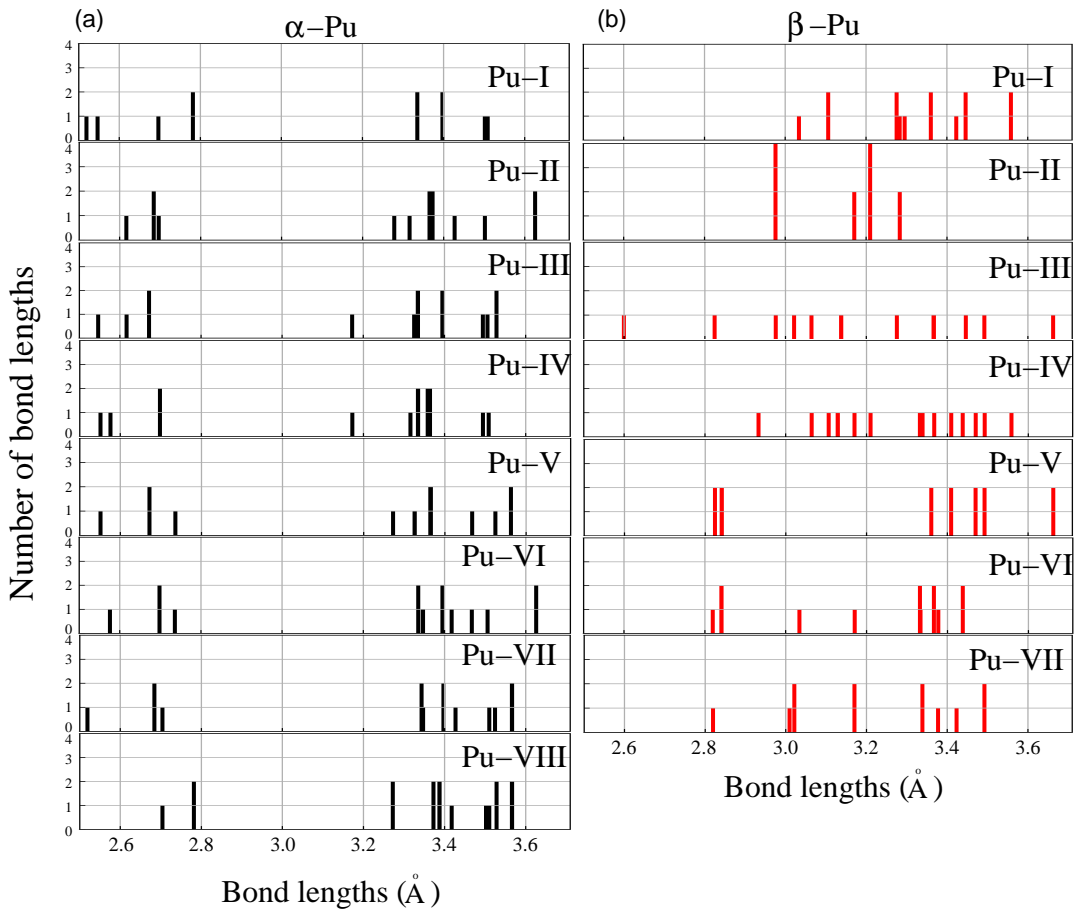}
 \caption{Distribution of bond lengths of each Pu-site in the (a) $\alpha$ and (b) $\beta$ phases.}
\label{fig:NN_dist_lowSymm}
\end{figure*}

From the structural point of view, the distribution of bond lengths (Fig.~\ref{fig:NN_dist_lowSymm}) within these two phases suggest  different local  environments for the Pu-5$f$ states  within the unit cells.  In our analysis, we considered two groups of bond lengths, one corresponding to short bonds (2.50 - 3. 12 \AA{}) and the other to long bond lengths (3.13  - 3.71 \AA{}). Similar classifications were used in previous reports.~\cite{ellinger, willsNatCom}  In Fig.~\ref{fig:NN_dist_lowSymm}(a) we display the distribution of bond lengths corresponding to the $\alpha$ phase.
As  pointed out by Zhu et al.,~\cite{willsNatCom} the distribution of bond lengths in this phase shows three groups of plutonium sites, and two well defined groups of short and long bond lengths. The Pu-I site is the one with the largest number of short bonds, \textit{i.e}, it has five short bonds and seven long bonds. As presented in  table I, the average short bond length of this site is 2.67 \AA{}. On the other hand, Pu-VIII site has only three short bonds , with average bond length of 2.76 \AA{}.  We have an intermediate situation for the sites from Pu-II to Pu-VII.
In Fig.~\ref{fig:NN_dist_lowSymm}(b) we show the same distribution for the $\beta$ phase.  Although, the bond distribution is more spread in this case, we observe the existence of similar Pu sites within the unit cell of $\beta$-Pu. In fact, as discussed by us in our previous work~\cite{whb_betapu}, the Pu-III site has  five short bond lengths, with average of 2.90 \AA{}, whereas Pu-I has only three short bonds.  
Therefore, one can expect a larger hybridization concerning Pu-5$f$ states in Pu-I (Pu-III) site in $\alpha$-Pu ($\beta$-Pu) whereas these states should be more localized (less hybridized) in the Pu-VIII (Pu-I) site. 

\begin{table}[!htb]
\label{NN_res} 
\caption{Number of short and long bond lengths in $\alpha$ and $\beta$-Pu. \={d} denotes the weighted average bond lengths within each subgroup.}
\begin{ruledtabular}
\begin{tabular}{ccccc}
Pu site & short  & \={d}$_{short}$ (\AA{}) & long & \={d}$_{long}$(\AA{}) \\ \hline
             &              &   $\alpha$-Pu              &            &        \\ 
$I$     &  5  & 2.67 & 7 & 3.43 \\
$II$    &  4  & 2.67 & 10  & 3.43 \\
$III$   &  4   & 2.63 & 10 & 3.40   \\
$IV$    &  4  & 2.63 & 10 & 3.36 \\
$V$     &  4   & 2.66 & 10 & 3.42  \\
$VI$    &  4   & 2.68  & 10 & 3.45 \\
$VII$   &  4   & 2.65 & 10  & 3.44 \\ 
$VIII$  &  3 &  2.76    & 13 & 3.44 \\ \hline 
            &              &   $\beta$-Pu              &            &        \\ 
$I$     &  3   & 3.08 &  11 & 3.39 \\
$II$    &  4   & 2.98 &  8  & 3.22 \\
$III$   &  5   & 2.90 & 7 & 3.39   \\
$IV$    &  4   & 3.06 & 10 & 3.38 \\
$V$     &  4   & 2.83 & 10 & 3.48  \\
$VI$    &  4    & 2.88  & 9 & 3.35 \\
$VII$   &  4   & 2.97 &  8 & 3.35 \\  \hline
$\gamma$-Pu & 4   & 3.03 &  6 & 3.25 \\
$\delta$-Pu  &  -- & -- & 12 & 3.29 \\
$\delta'$-Pu & --  & -- &  12 & 3.27 \\
$\epsilon$-Pu & --  & -- &  14 & 3.36 \\ 
\end{tabular}
\end{ruledtabular}
\end{table}

\begin{figure*}
 \includegraphics[scale=0.35]{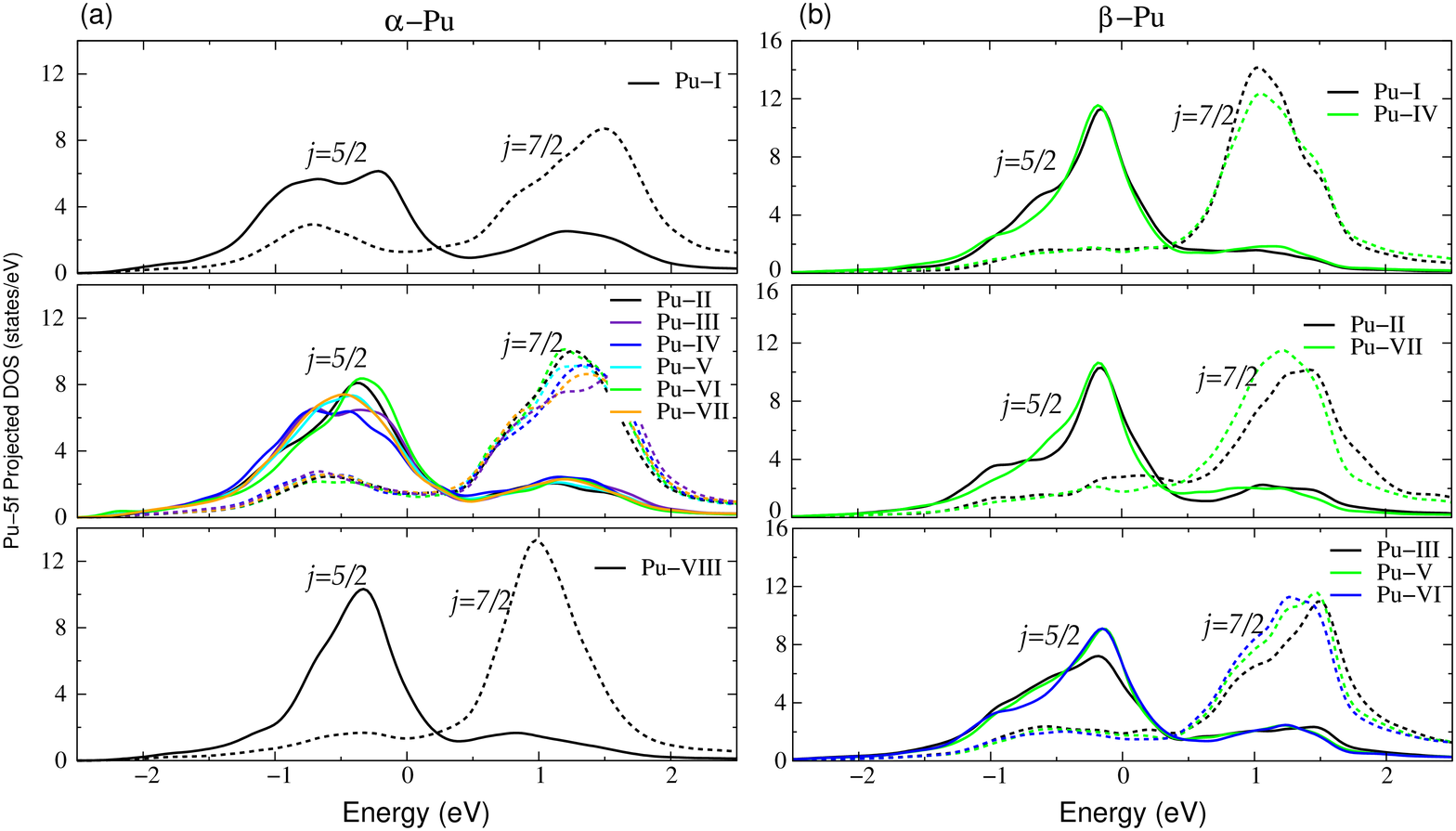}
 \caption{DFT(GGA) Pu-5$f$ projected density of states for the different Pu sites in $\alpha$ and $\beta$-Pu. Continuous (dashed) lines denote the Pu-5$f_{5/2}$ (Pu-5$f_{7/2}$) projected density of states.}
 \label{fig:pdos_DFT_lS}
\end{figure*}

Our DFT calculations show that some effects of the distinct crystallographic sites can be noticed in the Pu-5$f$ projected density of states, shown in Fig.~\ref{fig:pdos_DFT_lS}. As can be noticed for the $\alpha$-Pu, the Pu-5$f_{5/2}$ states are more broader in Pu-I site than in Pu-VIII site. In the case of $\beta$-Pu, there similar crystallographic sites give rise to similar projected density of states, such as in projected density of states of sites I and IV.

\begin{figure*}
 \includegraphics[scale=0.4]{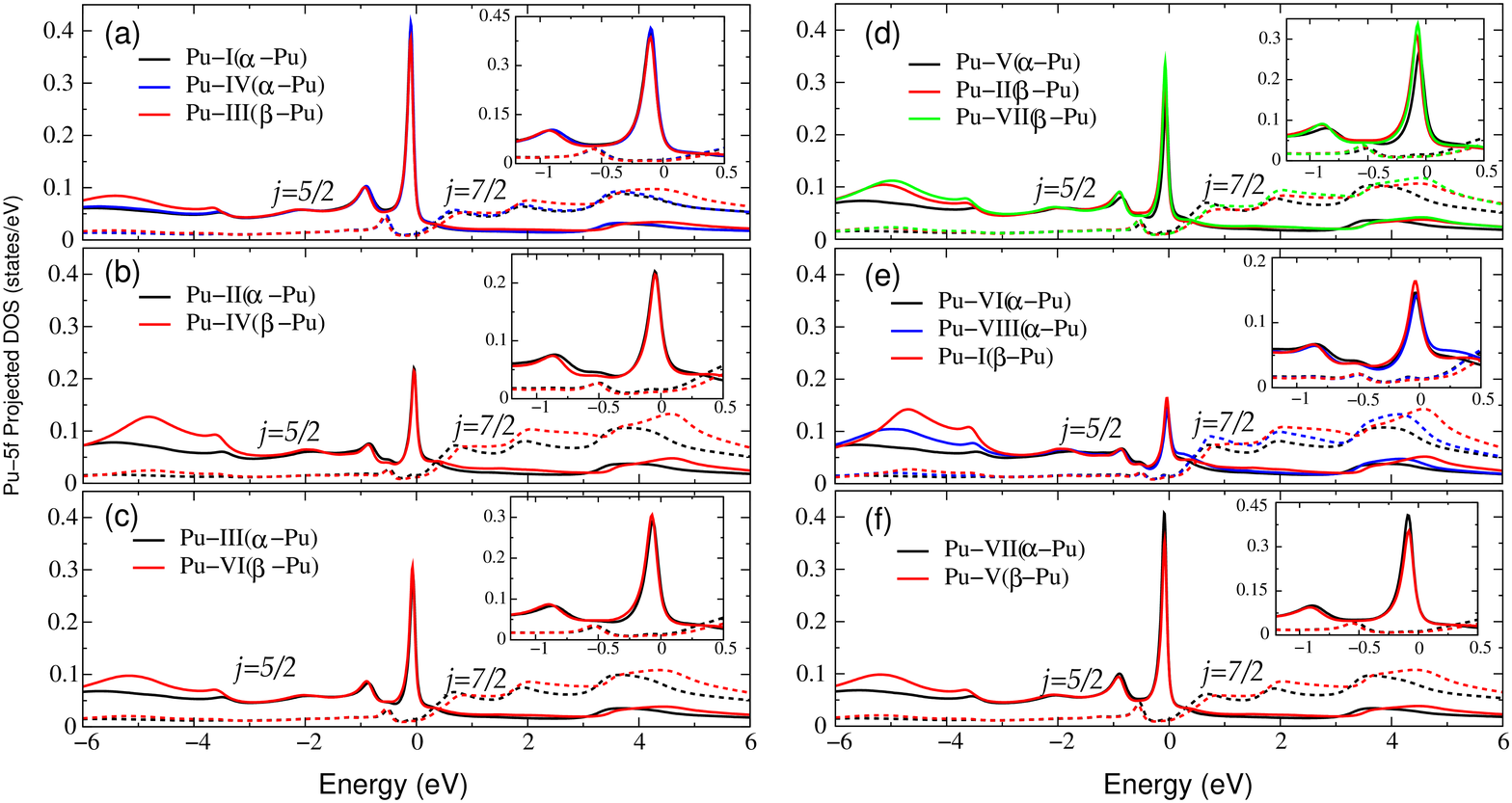}
 \caption{DFT+DMFT(OCA)  Pu-5$f$ projected density of states at 500 K for the different Pu sites in $\alpha$ and $\beta$-Pu. Continuous (dashed) lines denote the Pu-5$f_{5/2}$ (Pu-5$f_{7/2}$) projected density of states. Insets display the projected density of states in a smaller energy window (-1.2 eV $\leq$ E$_{F}$ $ \leq$ 0.5 eV).}
 \label{fig:pdos_DMFT_lS_OCA}
\end{figure*}

Next, we present the Pu-5$f$ projected density of states (Fig.~\ref{fig:pdos_DMFT_lS_OCA})  of each site in $\alpha$ and $\beta$-Pu at 500 K. As found for the high-symmetry Pu phases,  the electronic correlations taken into account within DMFT(OCA), give rise to strongly renormalized quasiparticle peaks close to the Fermi energy and Hubbard bands at higher energies. There are also multiplets in spectra of Pu-5$f_{5/2}$ and Pu-5$f_{7/2}$ characters below the main quasiparticle peak.  The intensity of this quasiparticle peak is strongly site-dependent, as can be observed in Figs.~\ref{fig:pdos_DMFT_lS_OCA}(a) and (e). In fact, the sites where the Pu-5$f$ electrons have a strong  hybridization with the conduction electrons have a more intense quasiparticle peak and Kondo temperature. As discussed previously, the Pu-5$f$ electrons at Pu-I site of $\alpha$-Pu  and Pu-III site of $\beta$-Pu are more hybridized and therefore have a larger coherence Kondo scale. In contrast, the Pu-5$f$ electrons are more localized at the Pu-VIII site in $\alpha$-Pu and at Pu-I in $\beta$-Pu, which explains the less intense quasiparticle peak.  These findings are in good agreement with previous reports on site-dependent correlations in $\alpha$ and $\beta$-Pu.~\cite{willsNatCom,whb_betapu} Another common feature with the spectra of the high-symmetry Pu phases is the incoherent character of Pu-5$f_{7/2}$ states, which again are essentially gapped at  the Fermi energy. Overall, our findings show that Pu-5$f_{7/2}$ spectral function are incoherent, gapped at the Fermi energy, and essentially independent of the crystal structure.

\begin{figure}
 \includegraphics[scale=0.4]{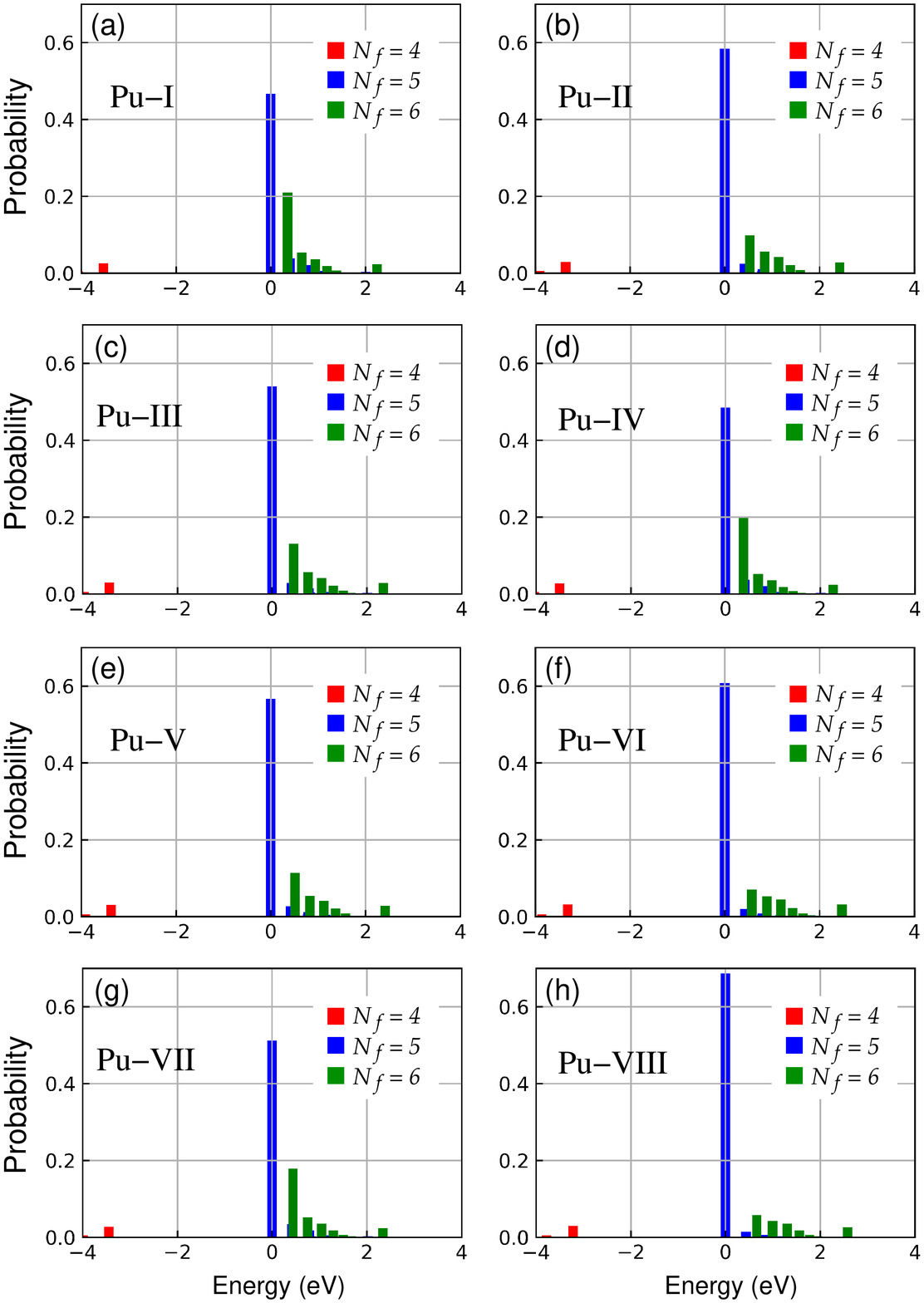}
 \caption{Valence histogram of Pu-5$f$ states associated with (a) Pu-I, (b) Pu-II, (c) Pu-III, (d) Pu-IV, (e) Pu-V, (f) Pu-VI, (g) Pu-VII, and (h) Pu-VIII sites of $\alpha$-Pu at 500 K.  N$_f$ denotes the number of electrons in the Pu-5$f$ manifold.}
 \label{fig:histo_lS_alpha_500K}
\end{figure}

\begin{figure}
 \includegraphics[scale=0.4]{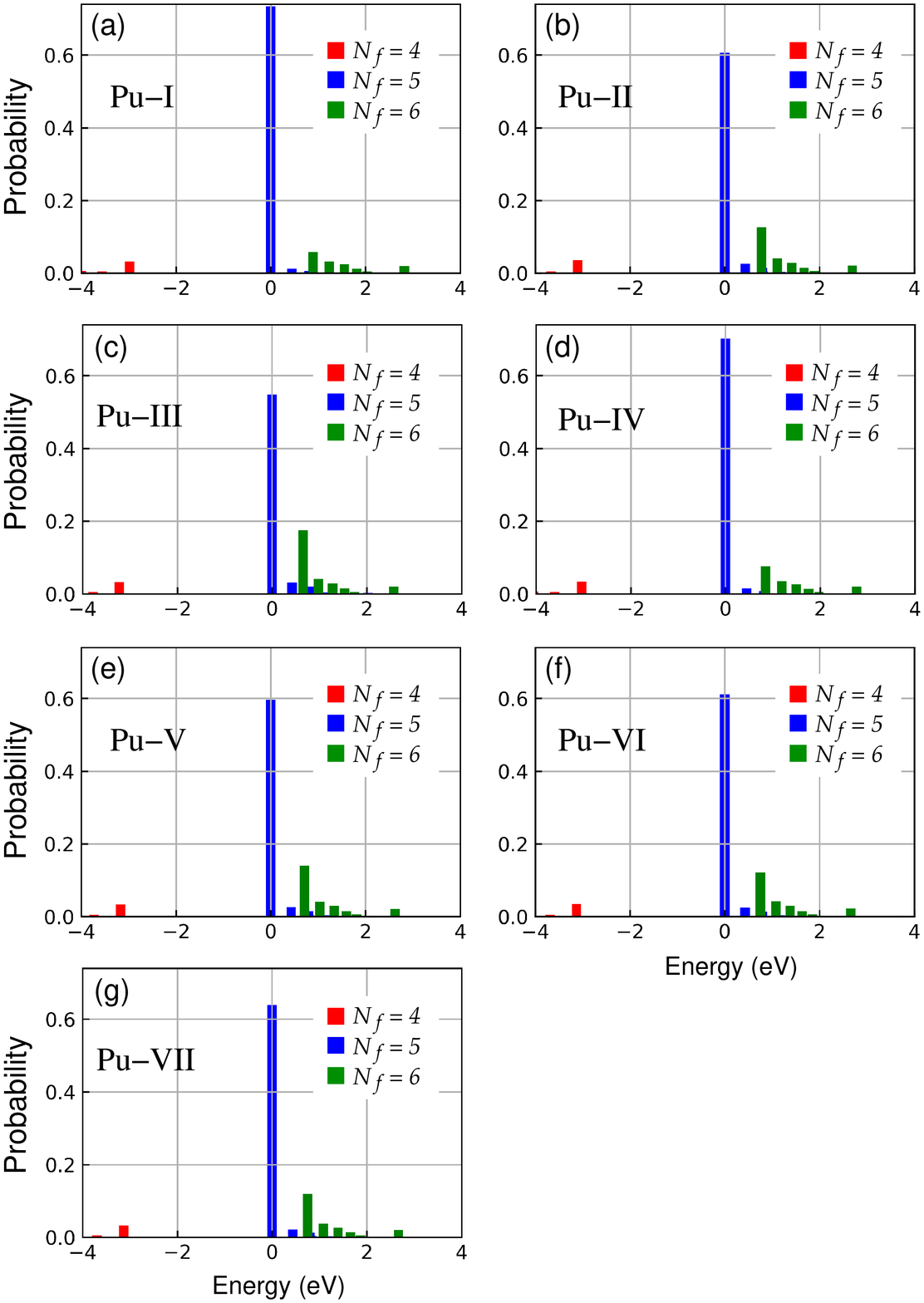}
 \caption{Valence histogram of Pu-5$f$ states associated with (a) Pu-I, (b) Pu-II, (c) Pu-III, (d) Pu-IV, (e) Pu-V, (f) Pu-VI, and (g) Pu-VII of $\beta$-Pu at 500 K.  N$_f$ denotes the number of electrons in the Pu-5$f$ manifold.}
 \label{fig:histo_lS_beta_500K}
\end{figure}

The mixed valent state is also observed for each site within $\alpha$ and $\beta$ phases. In Figs.~\ref{fig:histo_lS_alpha_500K} and \ref{fig:histo_lS_beta_500K} we show the valence histograms of each Pu site in $\alpha$ and $\beta$-Pu, respectively. We observe  that the 5$f^{5}$ configuration has a larger probability  in sites where the Pu-5$f$ are less hybridized, such as in Pu-VIII($\alpha$) and Pu-I ($\beta$) sites. On the other hand, fluctuations between 5$f^{5}$ and 5$f^{6}$ configurations are enhanced in sites where the Pu-5$f$ electrons are more hybridized, such as in Pu-I ($\alpha$) and in Pu-III ($\beta$) sites. These results emphasize the existence of site-dependent correlations in the low-symmetry phases of elemental Pu. In addition, the $\alpha$-Pu has some sites where the 5$f^{5}$ configuration is less probable in comparison to the high-symmetry phases and to the $\beta$ phase as well. In fact, the $\beta$-Pu has a mixed valence state which is more similar to the high-symmetry phases.

\subsection{Degree of electronic correlations, x-ray branching ratios, and orbital differentiation}

In order to quantify the degree of electronic correlations in the distinct phase of Pu we have examined our DMFT(OCA) self-energies. Although the vertex-corrected one crossing approximation is known to overestimate the renormalizations in 5$f$ systems,~\cite{zhu,whb_betapu} one can compare the renormalizations throughout the Pu allotropes and on the different Pu-5$f$ orbitals.  From the real part of the self-energies we evaluated the quasiparticle weights associated with Pu-5$f_{5/2}$ and Pu-5$f_{7/2}$ states, which are  given by
\begin{equation}
\label{eq:qwZ}
Z_{i}^{S} = \frac{1}{1-\partial_{\omega} Re \Sigma_{i}^{S}(\omega)|_{\omega = 0}}, 
\end{equation}
where $\Sigma_{i}(\omega)$ is the DMFT self-energy on real-frequency axis of $i$ state, where $ i = \{j=5/2, j=7/2 \}$ and $S$ denotes the inequivalent Pu-sites in the low-symmetry phases.

Our results are presented in Table II. First we point out that the occupancies are very similar for all phases, with essentially four electrons in Pu-5$f_{5/2}$ states and one in the  Pu-5$f_{7/2}$. This leads to very similar x-ray branching ratios which are useful quantities regarding the interpretation of X-ray absorption experiments. These ratios are evaluated using the Pu-5$f$ occupancies as follows~\cite{shim_xray_ratio,shimOCAPu} 
\begin{equation}
 B = \frac{3}{5} - \frac{4}{15}\left(\frac{1}{14-n_{5/2}-n_{7/2}}\right)\left(\frac{3}{2}n_{7/2}-2n_{5/2}\right).
\end{equation}
From table II one can observe that our calculated branching ratios are almost the same throughout the six phases of Pu, {\it i.e.} B($\alpha$) = 0.793, B($\beta$) = 0.796, B($\gamma$) = 0.794, B($\delta$) = 0.795, B($\delta'$) = 0.803, and B($\epsilon$) = 0.797. Overall, our findings are underestimated in comparison with the experimental values of 0.842 and 0.847, for $\alpha$ and $\delta$-Pu, respectively.~\cite{bratios_exp_PRBMoore} Moreover, our findings suggest that the difference of Pu-5$f$ occupancies across Pu phases is almost negligible and hard to detect using x-ray absorption from the core 4$d$ to the 5$f$ states.

Among the high-symmetry Pu phases, one observe that the Pu-5$f_{5/2}$ states are more renormalized in the $\delta$ phase, where $Z_{5/2}$ = 0.05. Further, we find that $Z_{5/2}(\gamma) < Z_{5/2}(\epsilon) < Z_{5/2}(\delta')$.  This indicates that electrons in Pu-5$f_{5/2}$ states are more correlated in the $\delta$ phase and more itinerant in the $\delta'$ phase. Although the average bond lengths in $\delta'$ are slightly smaller than in $\delta$-Pu, the corresponding valence histograms (see Fig.~\ref{fig:histo_DMFT_hs_OCA}) indicate larger valence fluctuations in the $\delta'$ phase, whereas in $\delta$-Pu the total occupancy is close to 5.
This feature can also be noticed in intensities of the quasiparticle peaks shown in Fig.~\ref{fig:pdos_DMFT_hs_OCA}. One can see that the quasiparticle peak is more intense in $\delta'$ than in $\delta$ phase, which suggest that the Pu-5$f$ electrons in the former are more hybridized while the same electrons in the latter are more localized. These findings also indicate that the Kondo energy scale is lattice dependent, with $T_{K,5/2}(\delta') > T_{K,5/2}(\epsilon) > T_{K,5/2}(\gamma) > T_{K,5/2}(\delta)$. 

Concerning the low-symmetry phases, we find that the quasiparticle weights are site dependent. In the $\alpha$-Pu the Pu-I site, where the Pu-5$f$ electrons are more hybridized, $Z_{5/2}$ is twice as large as the same in Pu-VIII site, where the electrons are more localized. The same trend is also seen in the $\beta$-Pu, where the $Z_{5/2}$ are smaller at the Pu-I and Pu-IV sites where the electrons are weakly hybridized. Furthermore, our results suggest that electrons in Pu-5$f_{5/2}$ states at sites in Pu-VIII (Pu-I) of $\alpha$-Pu ($\beta$-Pu) crystal structure are  correlated as in $\delta$-Pu. These  site dependent correlations point out the existence of multiple Kondo scales in these low-symmetry materials, as reported by previous theoretical studies.~\cite{willsNatCom,whb_betapu} 
In addition, we evaluate the electronic specific heat coefficients ($\gamma$), which are also shown in Table II.  $\delta$-Pu has the largest coefficient  among the high-symmetry phases, with $\gamma$ of around 297 mJmol$^{-1}$K$^{-2}$ which is overestimated in comparison to the experimental values of 43 - 64 mJmol$^{-1}$K$^{-2}$.~\cite{lashleyPRL,havelaPRB,lashley} Besides the similar quasiparticles weights,  $\gamma$-Pu has also a large specific  heat coefficient of $273.9$  mJmol$^{-1}$K$^{-2}$. It is important to mention that the other high-symmetry phases have much smaller coefficients, where $\delta'$-Pu has the minimum $\gamma$ of 114  mJmol$^{-1}$K$^{-2}$. Interestingly, the coefficient of Pu-I site in $\beta$-Pu is larger than $\gamma$ of $\delta$-Pu, while in $\alpha$ phase we find moderate values of $\gamma$. 

\begin{table}[!htb]
\label{Zns_res} 
\caption{Quasiparticle weights and occupancies of Pu-5$f_{5/2}$ and Pu-5$f_{7/2}$ states in $\gamma$, $\delta$, $\delta'$, $\epsilon$, $\alpha$, and  $\beta$-Pu at 500 K. The corresponding specific heat coefficients $\gamma$ (mJmol$^{-1}$K$^{-2}$) and x-ray branching ratios $B$ are also presented.}
\begin{ruledtabular}
\begin{tabular}{ccccccc}
Pu phase/site & $Z_{5/2}$ & $Z_{7/2}$ & $\gamma$ &$n_{5/2}$ & $n_{7/2}$ & $B$ \\ \hline
$\gamma$-Pu &  0.06  & 0.23   &  273.9 &  4.04 & 1.07 & 0.794\\
$\delta$-Pu       &   0.05  & 0.28   & 296.8  & 4.03  &  1.02 & 0.795 \\
$\delta'$-Pu      &   0.11    & 0.17   & 114.2  &  4.14 & 1.07 & 0.803\\
$\epsilon$-Pu  &  0.07    &  0.22  & 142.0  & 4.07  & 1.04 & 0.797\\ \hline 
        &           & $\alpha$-Pu  & &    &  &  \\  
$I$     &  0.10    &  0.11 & 118.4 & 4.12 & 1.24 & 0.797 \\
$II$    &  0.06    &  0.30 & 164.2 & 4.03 & 1.22 & 0.790\\
$III$   &  0.07    &  0.19 & 155.2 & 4.05 & 1.24 & 0.791\\
$IV$    &  0.10    &  0.12 & 106.1 & 4.12 & 1.23 & 0.797   \\
$V$     &  0.06    &  0.23 & 158.6 & 4.04 & 1.23 & 0.790 \\
$VI$    &  0.05   &  0.36 & 217.4 & 4.01 & 1.23 & 0.788 \\
$VII$   &  0.09    &  0.14 & 115.5 & 4.10 & 1.22 & 0.796 \\ 
$VIII$  & 0.05      &  0.36   &  208.6    & 4.03 & 1.15 & 0.792 \\
\hline 
       &           & $\beta$-Pu  & &    &   &  \\  
$I$     &  0.05    &  0.34 & 375.8 & 4.05 & 1.08 & 0.795 \\
$II$    &  0.08    &  0.17 & 252.7 & 4.07 & 1.15 & 0.795 \\
$III$   &  0.10    &  0.13 & 159.2 & 4.11 & 1.17 & 0.798 \\
$IV$    &  0.05    &  0.28 & 371.1 & 4.05 & 1.10 & 0.794 \\
$V$     &  0.08    &  0.17 & 219.9 & 4.09 & 1.15 & 0.796 \\
$VI$    &  0.07    &  0.20 & 241.8 & 4.07 & 1.15 & 0.795 \\
$VII$   &  0.07    &  0.20 & 248.9 & 4.08 & 1.12 & 0.796 \\ 
\end{tabular}
\end{ruledtabular}
\end{table}

\begin{figure*}[!htb]
 \includegraphics[scale=0.4]{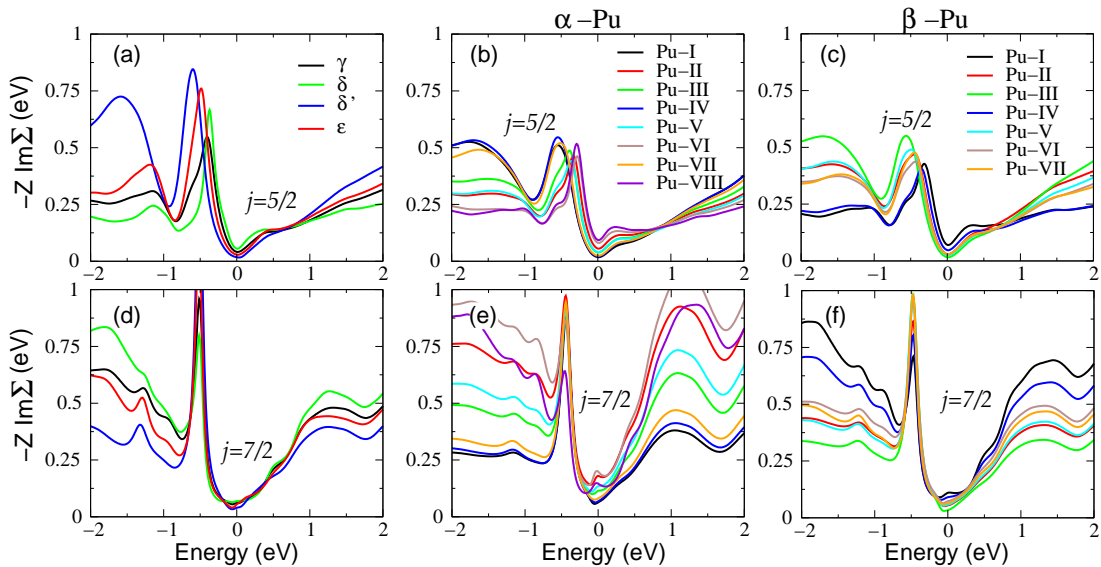}
 \caption{Renormalized imaginary part of self-energies of  Pu-5$f_{5/2}$ (upper column) and Pu-5$f_{7/2}$ (lower column)  at 500 K for the distinct phases and Pu-sites.  (a) and (d) for $\gamma$, $\delta$, $\delta'$ and $\epsilon$-Pu. In (b) and (e) we have the renormalized self-energies for each Pu-site of $\alpha$-Pu, whereas in (c) and (f) the same for $\beta$-Pu. }
 \label{fig:Zsig_Puphases}
\end{figure*}

Although the Pu-5$f_{7/2}$ states show larger quasiparticle weights, the spectral function of those states are suppressed and very incoherent as shown in Figs.~\ref{fig:pdos_DMFT_hs_OCA} and \ref{fig:pdos_DMFT_lS_OCA}. This suggests that $Z_{7/2}$ as evaluated according to Eq.~\ref{eq:qwZ}, can not be used to quantify the degree of coherence of the Pu-5$f_{7/2}$ states in elemental Pu. In Fig.~\ref{fig:Zsig_Puphases} we show the renormalized imaginary components of the self-energies of all Pu phases.  As one can see, both Pu-5$f_{5/2}$ and Pu-5$f_{7/2}$ states have small scattering rates near the Fermi energy,  where the quasiparticle peak of Pu-5$f_{5/2}$ character appears. The largest contribution in this case comes from the Pu-VIII site in $\alpha$-Pu, where the Pu-5$f$ are more localized.  On the other hand, the renormalized imaginary part of 5$f_{7/2}$ self-energies are very intense at higher energies, where the Pu-5$f_{7/2}$ states are located. As a result, there is a strong suppression of the Pu-5$f_{7/2}$ spectral function.  The $\alpha$-Pu is the phase where this renormalized imaginary-part of Pu-5$f_{7/2}$ is more intense (see Fig.~\ref{fig:Zsig_Puphases}(e)). In $\beta$-Pu, the Pu-I site self-energy leads to the highest suppression, followed by Pu-IV site.

Therefore, our results demonstrate that electronic correlations in elemental Pu are very sensitive to the 5$f$ occupancy, which is a clear fingerprint of strong correlations. Overall, the sites where 5$f$ occupancy is close to $5$, {\it i.e.} where the $f^5$ configuration has higher probabilities, have more correlated Pu-5$f_{5/2}$ states. The Pu-5$f$ occupancy, in turn, is ruled by the local crystallographic environment of each Pu atom within the crystal structure. We also find that orbital differentiation is a common feature within all the six phases of elemental Pu. The orbital differentiation can be explained based on the difference between the renormalized imaginary components of Pu-5$f_{5/2}$ and Pu-5$f_{7/2}$ self-energies, where the latter is strongly enhanced at higher energies.

\section{Conclusions}

In summary, we have performed DFT+DMFT(OCA) calculations to investigate the electronic correlations in all six phases of elemental Pu.  Our calculations indicate the presence of strong renormalizations in the Pu-5$f$ spectral function for all phases, with quasiparticle peaks near the Fermi energy, Hubbard bands at higher energies, and quasiparticle multiplets which are common features of Hund's -Racah metals. Overall, the Pu-5$f_{5/2}$ states are more correlated in sites where the occupancy of these states are close to $5$. More important, we find that orbital differentiation is a common feature throughout the six phases of elemental Pu, where the Pu-5$f_{5/2}$ and Pu-5$f_{7/2}$ states have very distinct spectral functions. This orbital differentiation  can be understood from the renormalized self-energies which are strong orbital dependent.

We thus arrive to a physical picture of Plutonium which is quite different from the early views which surmised that the low symmetry dense phases were weakly correlated while the high symmetry low density phases were anomalous. Instead we find that {\it all} the  Pu phases are strongly correlated and the strength of correlations evolve in a subtle way from site to site and among orbitals as the result of the interplay of Coulomb interactions and crystallographic environment. This new paradigm will be very useful in future studies of this material, and other strongly correlated 5$f$ electron materials.

\section{Acknowledgments}
This work was supported by the U.S. Department of Energy, Office of Science, Basic Energy Sciences as a part of the Computational Materials Science Program. The authors are grateful to Ran Adler for many useful discussions.

\end{document}